# Geometric and electronic properties on stage-1 FeCl$_3$-graphite intercalation compounds

Wei-Bang Li[1], Shih-Yang Lin[2], Ming-Shuei Tsai[1], Ming-Fa Lin[1,3], Kuang-I Lin[4]

[1]Department of Physics, National Cheng Kung University, Tainan, Taiwan

[2]Department of Physics, National Chung Cheng University, Chiayi, Taiwan

[3]Hierarchical Green-Energy Materials (Hi-GEM) Research Center, National Cheng Kung University, Tainan, Taiwan

[4]Center for Micro/Nano Science and Technology, National Cheng Kung University, Tainan, Taiwan

Email: weibang1108@gmail.com (W.B. Li), mflin@ncku.edu.tw (M. F. Lin), kilin@mail.ncku.edu.tw (K. I. Lin)

**Abstract**

The calculated results of FeCl3 graphite intercalation compounds show the detailed features. The stage-1 FeCl3-graphite intercalation compounds present diversified electronic properties due to the intercalant. The first-principles calculations on VASP are utilized to analyze the essential properties, such as the geometric structures, spatial charge distributions, charge variations, band structures and density of states. The density of states displays full information for an explanation of the hybridizations with the special structures van Hove singularities on it. The van Hove singularities in graphite-related systems are very important and can provide full information for examining the intercalation effects. The orbital-decomposed density of states for C atoms shows that the π bondings are orthogonal to the sp2 bondings and the C-C bondings retain in the intralayer C atoms. The Fe atoms and Cl atoms form the Fe-Cl bondings at some unique energy range, presenting the multi-orbital hybridizations of [4s, $3d_{xy}$, $3d_{yz}$, $3d_{xz}$, $3d_{x^2-y^2}$, $3d_{z^2}$]-[$3p_x$, $3p_y$, $3p_z$]. For C-Cl and Cl-Cl bonds, the unique van Hove singularities exhibit their coupling interactions, revealing the multi-orbital hybridizations of [$3p_x$, $3p_y$, $3p_z$]-[$3p_x$, $3p_y$, $3p_z$] and [3s, $3p_x$, $3p_y$, $3p_z$]-[3s, $3p_x$, $3p_y$, $3p_z$], respectively. The Fe-Cl bondings arise from multi-orbital hybridizations of [4s, $3d_{xy}$, $3d_{yz}$, $3d_{xz}$, $3d_{x^2-y^2}$, $3d_{z^2}$]-[$3p_x$, $3p_y$, $3p_z$]. Due to the band structures and density of states, the multi-orbital interactions between intercalants and graphene layers dominate in the low-lying energy range. The charge transfers per atom (electrons/atom) for C, Fe, Cl are -0.02 e/atom, -0.28 e/atom and +0.46 e/atom, respectively. Thus, the C atoms in graphene layers present as positive ones after intercalation, i.e., the graphite system exhibit p-type doping features in agreement with previous study[1].

*Introduction*

A bulk graphite[2-5], which is composed of multi-layer graphene systems, has attracted many theoretical and experimental researches, e.g. basic science, engineering and applications. In general, such condensed-matter systems become *n*-type or *p*-type metals depending on the species of interacting matter. For instance, the intercalation of an alkali metal into graphite will turn this system into an *n*-type metal because of free conduction electrons, whereas the intercalation of FeCl$_3$ turns it into a *p*-type metal because of free conduction holes. The interesting features of geometric structures and electronic properties lead to the application technology, such as green energy, and Lithium ion batteries.

Lithium ion batteries (LIB) are well-known rechargeable batteries and promising for various devices[6], e.g. cell phones, vehicles and almost all portable equipment, because of the reversible intercalation/de-intercalation of Li+ into/from the host lattices of either graphite-layered anodes or other layer-based cathodes. The intercalation/de-intercalation features on graphite are very important for the development of LIBs. Recently, many research papers mainly focused on the distinguished guest atoms/ions like Li, Na, K[7,8] or ion clusters like $AlCl_4$, $Al_2Cl_7$ [9-11]; also, battery capacity was gained with ferric chloride graphite intercalation compounds ($FeCl_3$-GICs) [12,13]. The $FeCl_3$-GIC is found to possess a high reversible capacity, cycling stability and an outstanding rate, demonstrating its potential as an electrode material in LIB. $FeCl_3$ is known to extract electrons from graphite, thereby creating a *p*-type doping effect[14].

In experiments, researchers synthesized and exfoliated the $FeCl_3$-GICs by mixing the $FeCl_3$ and graphite under proper temperature in dimethyl formamide (DMF) solvent[12]. These studies monitored the evolution of graphite in intercalation and exfoliation processes by X-ray diffraction (XRD) with Cu Kα radiation and a Renishaw inVia Raman microscope, and then analyzed the results on a Zeiss Supra 55 field-emission scanning electron microscope (SEM), FEI Tecnai G220 high-resolution transmission electron microscope (TEM), and Bruker multimode 8 atomic force microscope (AFM) under scan Asyst mode. Taking the stage-1 $FeCl_3$-GICs as an example, the pristine structures of graphite are broken with the expanded interlayer spacing. According to the results, the stage-1 structure of $FeCl_3$-GICs is explored by XRD. For pristine graphite, the periodic interlayer distance between each layer is about 0.33 nm; after intercalation with $FeCl_3$, the interlayer distance is significantly increased to 0.96 nm. The $FeCl_3$-GIC results in a larger capacity than pure graphite. Also, the $FeCl_3$-GIC exhibits a better cycling stability and rate capability compared with pure $FeCl_3$[13]. Other experiments even show that few-layered $FeCl_3$-GIC has a better rate capacity than stage-1 $FeCl_3$-GIC, which may be due to the residual $FeCl_3$ in few-layered $FeCl_3$-GIC as the main active material; Li-ions rapidly insert themselves into or extract from the interlayered spacing due to the nanoscale thickness and porous structure.

As an electrode for Li-ion batteries, stage-1 $FeCl_3$-GIC possesses reversible capacities of 525 mAhg$^{-1}$ after 100 charge-discharge cycles[13]. This is considerably higher than the capacity for expanded graphite, which is 113 mAhg$^{-1}$. Moreover, the in-plane conductivity at room temperature for stage-1 $FeCl_3$-GIC is $1.1\times10^5$ Ωcm$^{-1}$, which is higher than the $2.5\times10^4$ Ωcm$^{-1}$ in graphite. Not only that, but the stage-2 $FeCl_3$-GIC also exhibits a much higher reversible capacity with a specific capacity of 719 mAhg$^{-1}$ after 100 charge-discharge cycles, and even 910 mAhg$^{-1}$ after 170 charge-discharge cycles. The special and important performance of $FeCl_3$-GICs undoubtedly results from the unique geometric structure of intercalations and the stage index; in other words, the interlayer spacings of graphite are enlarged because of the intercalations of FeCl$_3$-GICs, and the enlarged spacings provide larger channels for the electrochemical reaction while promoting fast Li-ion transport. Although the reversible cycle capacity of stage-1 FeCl$_3$-GIC is lower than that of stage-2 FeCl$_3$-GIC, it is still a significant improvement and worthy of further exploration. Our work is focused mainly on the study of stage-1 FeCl$_3$-GIC, and we will show details of the essential properties of electrons, e.g. geometric structure/band structure/density of state/ spatial charge distribution/charge variation.

*Theoretical calculation*

Up to now, first-principles calculations[15] are the dominant method for solving the complicated interactions through some approximations and simplifications to obtain the eigenvalues and eigenfunctions; going a step further, we can also model and explore the essential properties. The Vienna Ab-initio Simulation Package (VASP)[16] computes the approximate solutions within density functional theory (DFT)[17-20] to solve the Kohn-Sham equations. The exchange-correlation energy due to the electron-electron interactions is calculated from the Perdew-Burke-Ernzerhof frunctional (PBE) [21-23] under generalized gradient approximation (GGA); projector-augmented wave methods (PAW)[24,25] are utilized to evaluate the electron-ion interactions. A primitive unit cell depends on the geometric distributions of host and guest atoms. In our calculations, the Brillouin zone is set to be 20×20×20 and 50×50×50 k point meshes for geometric structures and electronic distributions.

*Crystal symmetry in a large primitive unit cell*

In pristine AA-stacking graphite, each layer is bound by the weak but significant van der Waals interactions due to the perpendicular p$_z$ orbital between the neighboring layers. The strong $\sigma$ bondings appear between carbon atoms in hexagonal structures on every plane, and the bond length is 1.42 Å. In AA-stacking configurations, the interlayer distance of each neighboring layer is rouhly 3.55 Å [26,27], which is large enough to be intercalated/de-intercalated by certain atoms, ions, or even layered materials. Up to now, the graphite intercalation compounds of alkali metal atoms, e.g. Li, Na, K and Rb, are well-know as are some molecular clusters, e.g. AlCl$_4$, N$_2$O$_5$, Na(ethylenediamine)C$_5$[28]. The interlayer distances between neighboring carbon layers will depend on the species of intercalations, for instance, the distances are 3.76 Å and 5.65 Å for, respectively, Li atoms and Rb atoms[8]. Apparently, the interlayer distances are easily affected by intercalations. However the bond lengths between carbons in hexagonal structures, or in the same planes, are only slightly changed from 1.42 Å to 1.40 Å. The planar hexagonal structure is retained, so that the $\sigma$ bondings due to the (2s, 2P$_x$, 2P$_y$) orbitals of the carbon atoms seldom take part in the orbital hybridizations between intercalated atoms/ions/clusters and carbon atoms.

The optimized structure for the FeCl$_3$-graphite intercalation compound is shown in Figure 1. This supercell with lattice constants a=b=12.12 Å, which is constructed by combining 2×2 unit cells of

FeCl$_3$ and 5×5 unit cells of graphene as mentioned in previous works[29,30], covers 50 carbon atoms, 8 iron atoms and 24 chlorine atoms, due to the high symmetry, and is quite big. In addition, the FeCl$_3$ also forms layered structures between the carbon layers. The large spacings of interlayer distances between graphenes allow the guest FeCl$_3$ to be intercalated. Calculation resultsshow that the optimized interlayer distances between the pristine carbon layers are enlarged from 3.55 Å to 9.200 Å. The calculated results are very similar to the X-ray diffraction experimental observations, which showed an interlayer distance of 9.36 Å [12,29,31-33]. Interestingly, the tremendous changes of interlayer distance reveal that the interactions between interlayered 2P$_z$ orbitals become much weaker after intercalation. The pristine graphite has a very stable AB-stacking configurations, but it turns into AA-stacking; this is also the case for other stage-1 intercalations, e.g. alkali metal atoms.

*Electronic properties*
*1.Band structure*

Up to know, the usual stacking configurations of AAA, ABA and ABC have been synthesized in experiments. The AAA stacking is often presented in stage-1 graphite intercalation compounds because of the higher symmetry. However, natural graphite almost only takes on ABA and ABC structures. The three stackings of AAA, ABA and ABC are predicted to have unusual energy bands, respectively, and this indicates the important and essential properties of the special charge distributions and the ratio of free electrons/holes. It is well-known that the monolayer graphene is a semiconductor with zero gap, and there exist linearly intersecting valence and conduction bands at the Dirac-cone point. Thus the density of states vanishes at the Fermi level. The linear Dirac-cone structures on the k-space ($k_x$, $k_y$) planes are clearly revealed in the AAA-stacking and ABC-stacking graphites. In general, the energy of the Dirac point is quite sensitive to the $k_z$ component wave vector. Among the three kinds of stacking, the AAA-stacking graphite possesses a periodical arrangement of graphite layers in the z-direction and creates a band width as roughly 1 eV. Also, the density of state exists at the Fermi level. That is to say, the free electrons and holes appear simultaneously in the range of -0.5~0 eV and 0~0.5 eV, respectively. For ABA- and ABC-stacking graphites, the band widths in terms of the $k_z$-dependent π-electronic band are 0.2 eV and 0.03eV, respectively, near the Fermi level along KH line in $k_z$-dependent band structure. As a result, the AAA-, ABA- and ABC-stacking graphites are semimetalic with the free electrons and holes appearing simultaneously, while the latter two only possess relatively few free carriers. The electronic structures exhibit severe changes after the intercalation. That is to say, the electronic structures strongly depend on the properties of the intercalant materials. In this chapter, we intercalate FeCl$_3$ as an anion into graphitic layers.

It is well-known that the electrons situated on the outmost orbitals of an atom or molecule, i.e. the valence electrons, will dominate the size, electron binding energy and other chemical reactivity. Moreover, the valence electron of anions tends to be removed due to its relatively weak binding energy.

The negative charges in anions drive them to interact with surrounding molecules and ions. Also, the anions polarize the electronic clouds of nearby molecules in the opposite direction that cations do. The anions have a polarizability and thus tend to strongly interact more with surrounding molecules than the less polarizable neutral molecules and cations. In addition, the valence electron binding energy in anions is often smaller than neutral molecules and cations. In many case of intercalation, the intercalant species will critically govern the p- or n-type, as mentioned above. The system of graphite intercalation compounds exhibit n-type doping with the alkali-metal atom intercalants, for instance, the alkali-metal atoms will provide the electrons of the outmost orbitals, i.e., s-orbitals will lose electrons. And thus the graphite will behave like a metal due to the extra free electrons from the guest compound[11]. This result can be analyzed by the band structures and the density of states (DOS) as well. In the band structure, the Fermi level shifts from the Dirac point to the conduction points. The density of states allows a complete understanding of the relations among the different chemical bondings. The lowest valley in DOS with respect to the energy presents the blue shift of the Fermi level in agreement with the shift in band structure. Furthermore, the DOS indicates the hybridization in host-guest, host-host and guest-guest atoms. By the Bader analysis in VASP calculations, we can predict the charge transfers of the host and guest atoms. After intercalation, the alkali-metal atoms behave like cations due to the loss of electrons.

In our work, we explore the band structures, density of states (DOS), charge distributions and the differences between charge distributions of stage-1 $FeCl_3$-graphite intercalation compounds on VASP calculations with proper parameters. The abovementioned properties allow us to analyze and determine further essential mechanics, e.g. host-guest orbital hybridizations, the interlayer interactions, the intralayer interactions, and the intralayer σ- and π-bondings. For $FeCl_3$-graphite intercalation compounds, the first Brillouin zone in reciprocal lattice space is a symmetric hexagon, as exhibited in graphene. The 3D band structures of the hexagon are described along the high symmetry points in reciprocal lattice [Γ– K– M– Γ– A– H– L– A], as shown in Figure2.

As shown in Figure 3, the calculated band structures of $FeCl_3$-graphite intercalation compounds possess a Dirac cone with a blue shift of the Fermi level around 0.9 eV. The complicated horizontal bands mainly arise from the d-orbitals, i.e., the band structures of $FeCl_3$-graphite intercalation compounds are very similar to those of monolayer graphene, except for the d-orbitals. D. Zhan et. al published a research paper about $FeCl_3$-based few layer graphene intercalation compounds, which contains 2-layer- to 4-layer- intercalation compounds[29]. The band structures in the low-lying energy region is in agreement with our results. The interlayer distances between carbon layers are approximately 9.4 Å and 9.2 Å in $FeCl_3$-few-layer graphene intercalation compounds and $FeCl_3$-graphite intercalation compounds, respectively; the lattice constants of graphene and $FeCl_3$ are 6.06 Å and 2.46 Å, respectively. The rather large interlayer distance indicates that the feature of low-lying energy of this system is in good agreement with the monolayer graphene due to the loss of coupling interactions between the neighboring graphene layers.

The band structures at about 0 eV and 1.35 eV present d-orbitals of the iron atoms. The band structures excluding complicated d-orbitals are very similar to graphene, which possesses the Dirac points at 0 eV, with 0.9 eV with a red shift of the Fermi level. The red shifts indicate that there exist free holes in valence bands below the Dirac-cone point. That is to say, the system of stage-1 $FeCl_3$-graphite intercalation compounds turns into a *p*-type system. In addition, the charge transfers for our systems are examined and analyzed by Bader analyses in the VASP calculations. The charge transfers per atom (electrons/atom) for C, Fe, Cl are -0.02 e/atom, -0.28 e/atom and +0.46 e/atom, respectively. In our stage-1 $FeCl_3$-graphite intercalation compound unit cell, which contains 50 C atoms, 8 Fe atoms and 24 Cl atoms, the total transfers for C, Fe, Cl are -1 electrons, -2.24 electrons and +3.68 electrons, respectively. Thus, C atoms and Fe atoms present as positive ions, but the Cl atoms present as negative ions. Due to the above, the pristine Bernal graphite will possess more free holes after intercalations, i.e., it exhibits *p*-type doping features.

*2.Density of states*

Previous works showed that the graphite system can change into p- or n-type doping. During the intercalation, intercalants interact with the host material and transfer charges, which can be electrons or holes, to carbon atoms resulting in *p*- or *n*-type doping. As the diverse species of intercalants progress, the special orbital hybridizations come into play. The band structures, density of states, spatial charge distributions and the charge variations can give more information about the important chemical bondings, e.g., Fe-Cl bonds, Cl-Cl bonds, C-Cl bonds and C-C bonds.

In pristine graphite, the (2s, $2p_x$, $2p_y$) orbitals dominate the σ bondings in the planar honeycomb structure, which mainly form the geometric structures. The Van Hove singularities in 3D graphite-related systems are revealed as special structures. The atom- and orbital-decomposed density of states, as shown in Figure4, could provide full information and are useful in examining the host/guest effects after intercalations. In principle, these structures originate from the critical points in the energy-wave-vector space. In AB-stacking graphite, the parabolic dispersions have band-edge states at the M-point, in which they belong to the saddle points in the energy-wave-vector space. This indicates that π electronic states are accumulated near the M-point and will make important contributions to the essential physical properties. The saddle point is mainly owed to the interlayered interaction. In other word, the interactions between neighboring layer atoms will result in the asymmetry of valence and conduction bands. The low-lying $2p_z$ orbitals will be easily influenced after intercalation and even sensitive to the kinds of intercalants. On the other hand, the 2s, $2p_x$ and $2p_y$ orbitals can generate much deeper energy bands initiated by the Γ points at E= -3 eV.

In previous cases of diverse intercalations such as alkali-metals, e.g. Li/Na/K/Rb/Cs atoms, the weak but significant van der Waals interactions arise from the interlayer $2p_z$-$2p_z$ and $2p_z$-s orbital hybridizations in C-C and C-A bonds, respectively, where A represents alkali metals; these interactions also make the most important contribution to the low-lying π-electronic structures and thus dominate the essential physical properties. For instance, the number of stages will lead to quite different amounts of asymmetric free electrons and free holes; the distinct alkali metal atoms also cause much different interlayer distances because of their very different atom sizes.

In the case of $FeCl_3$ as an intercalant, this super big intercalant will have a charge distribution and charge variation in C-Cl bonds, as shown in Figure5. The charge variation reveals that the amount of difference between C and Cl atoms are not too much, and this agrees with the charge transfers mentioned above of only 0.02 electrons per atom from C to Cl atoms. This phenomenon probably indicates that the Cl atoms are close to Fe atoms and mainly form bonds with Fe atoms. Similarly, the charge distributions between intralayer neighboring C atoms are very strong due to the parallel $2p_z$ orbital hybridizations; after intercalations, the variation almost vanishes between the intralayer neighboring C atoms. That is to say, the C-C bonds in intralayer remains even intercalating. The anisotropic charge distributions between Fe atoms and Cl atoms reveal that there might exist interactions between them.

The van Hove singularities in graphite-related systems are very important and useful, being revealed as special structures and quite different from those in layered graphene systems. The atom- and orbital-decomposed density of states, as shown in Figure 4, could provide full information for examining the intercalation effects. In principle, these structures originate from the critical points in the energy-wave-vector space. The band structures are very complicated due to the d-orbitals below 0 eV and above 1.3 eV. However, the Dirac-cone point remains with a shift of Fermi level about 0.9 eV and is in good agreement with the lowest valley at E=~0.9 eV in the density of states. In Figure 4, the orbital-decomposed density of states for C atoms present that the σ bands are initiated from -2 eV and the difference between the initiation and Fermi level is roughly -2.9 eV ~ -3 eV centered from Fermi level, and it almost remains the same as pristine AB-stacking graphite. The orbital-decomposed density of states for C atoms also shows that the π bondings are orthogonal to the $sp^2$ bondings, and the π bondindgs are situated at – 1.8 eV without other singularities there. This means that there exist C-C bonds in the intralayer C atoms.

Based on the van Hove singularities in density of states, the Fe atoms and Cl atoms form the Fe-Cl bondings at – 4.7 eV ~ - 5 eV, - 3.2 eV ~ - 2.3 eV, - 1.5 eV ~ - 1.8 eV, - 1 eV ~ 0 eV and 1 eV ~ 2 eV, presenting the multi-orbital hybridizations of [4s, $3d_{xy}$, $3d_{yz}$, $3d_{xz}$, $3d_{x^2-y^2}$, $3d_{z^2}$]-[$3p_x$, $3p_y$, $3p_z$]. Corresponding to the band structures, apparently, the complicated d orbitals of Fe atoms, which originate from E < 0 eV and E > 1 eV, are the main orbitals of bonding with Cl atoms, especially -1 eV ~ 0 eV and 1 eV ~ 2 eV. As for C-Cl and Cl-Cl bonds, the unique van Hove singularities exhibit their couplings in - 6 eV ~ - 2 eV and roughly - 6.2 eV, respectively, revealing the multi-orbital hybridizations of [$2p_x$, $2p_y$, $2p_z$]-[$3p_x$, $3p_y$, $3p_z$] and [$3s$, $3p_x$, $3p_y$, $3p_z$]-[$3s$, $3p_x$, $3p_y$, $3p_z$]. The red shift of the Fermi level reveals the p-type doping effects of the intercalants. In the other word, the C-Cl bonds will affect the diversity of the van Hove singularities around Fermi level. Also, this results are in agreement with the spatial charge distributions and charge variations in the spacing between the host and guest layers. The low-energy physical properties of graphite and graphite intercalation compounds are dominated by the multi-orbital interactions between intercalants and graphene layers.

X-ray diffraction (XRD) is available for investigation of crystalline materials or graphene-related intercalation compounds[34-42]. The used wave length of the X-rays is close to the spacing between the atoms in lattice. The incident photons experience elastic scattering within the periodic structure of the lattice atoms and emerge as constructive or destructive patterns in some special directions. Graphite has typical XRD patterns in which the dominant peak relates to the basal planes (002). The scattering peak appears at an angle of $2\theta \approx 26.4°$ and corresponds to $d_{002}$=3.35 Å. The intercalation of stage-1 $FeCl_3$ increase the interlayer distance and thus the peak is displaced to new scattering angles at 9.1°, 18.6°, 28.1°, 37.8°, 50.4°. The interlayer distance $d$ of $FeCl_3$-graphite intercalation compounds is drastically increased to 9.6 Å, corresponding to the scanning electron microscope (SEM) measurements in previous studies. In addition, the characteristic Raman scattering is utilized for estimating the Fermi level of the graphite intercalation compounds.

Raman spectroscopy uses a monochromatic laser to interact with molecular vibrational modes and phonons in a sample. A shifting-down of the laser energy is called Stoke Raman scattering, while shifting-up is called Anti-Stoke Raman scattering through inelastic scattering. The peaks in Raman spectra exhibit some important information, e.g., the G peak (1580 cm$^{-1}$) comes from the C-C sp$^2$-orbital hybridizations, the D peak and the 2D peak present the first order scattering of the zone-boundary phonons and the second order scattering of the zone-boundary phonons, respectively. The G peak is also used to estimate the staging. In general, stage-1 graphite intercalation compounds possess a single G peak. For p-type graphite intercalation compounds:

$$E_T = E_L - \hbar\omega_{2D},$$

where $E_L$, $E_T$ and $\omega_{2D}$ represent the excitation energy, the energy corresponding to the electron-hole recombination process and the 2D frequency, respectively[43]. This is due to the ratio of 2D peaks and G peaks. The $E_T$ and $E_L$ are situated at the valence band and conduction band, respectively. Apparently, the equation is only applicable if $E_L$>2$E_F$. Applying a distinct wavelength of a laser allows us to detect the Fermi level. The previous experimental works put the Fermi level of 0.9 eV[31], which is in agreement with our calculations.

*Conclusion*

In summary, the stage-1 FeCl$_3$-graphite intercalation compounds present diversified electronic properties due to the intercalant. The first-principles calculations on VASP are utilized to analyze the essential properties, such as the geometric structures, spatial charge distributions, charge variations, band structures and density of states. The interlayer distances of pristine graphite are enlarged from 3.35 Å to 9.40 Å by the super big molecule FeCl$_3$, in which the unit cell consists of 50 C atoms, 8 Fe atoms and 24 Cl atoms. The density of states displays full information for an explanation of the hybridizations with the special structures van Hove singularities on it. The lowest valley in the density of states shifts from 0 eV to 0.9 eV. In other word, the Fermi level exhibits a red shift. Furthermore, the intercalant FeCl$_3$ results in a p-type doping. The interaction between interlayer C atoms becomes very weak but almost remain the same between the intralayer C atoms. The Fe-Cl bondings arise from multi-orbital hybridizations of [4s, 3d$_{xy}$, 3d$_{yz}$, 3d$_{xz}$, 3d$_{x^2-y^2}$, 3d$_{z^2}$]-[3p$_x$, 3p$_y$, 3p$_z$]. C-Fe bondings do apparently not exist in our results. In addition, for the FeCl$_3$-graphite intercalation compound, the large size and thickness hinder the transport of lithium ions and electrons. On the other hand, the layered structures with enlarged interlayer spacings allow the transport of lithium ions and electrons and provide additional sites for the accommodation of lithium, and thus exhibit a better storage performance as an electrode in lithium ion batteries. The differences of essential properties between distinct staging configurations are worth investigating as well. The stage-n systems with increasing n will possess decreasing concentrations and lower symmetries. Interestingly, for the FeCl$_3$-graphite intercalation compounds, the stage-1 case performs better than pure graphite in both theory and experiment, whereas the stage-2 case exhibits a higher reversible capacity and better stability than the stage-1 case[13]. The complicated mechanics of intercalation and de-intercalation in stage-n configurations, where n is greater than 2, are very important and worthwhile investigating further in the future in order to construct a theoretical framework coinciding with the presently performed experiments.

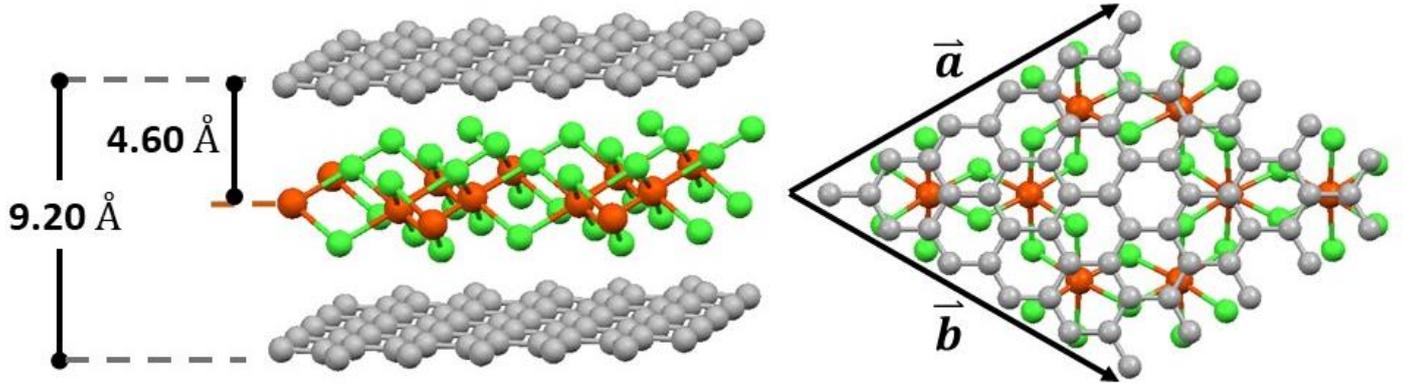

Figure 1 The geometric structure of FeCl$_3$-grapgite intercalation compound.

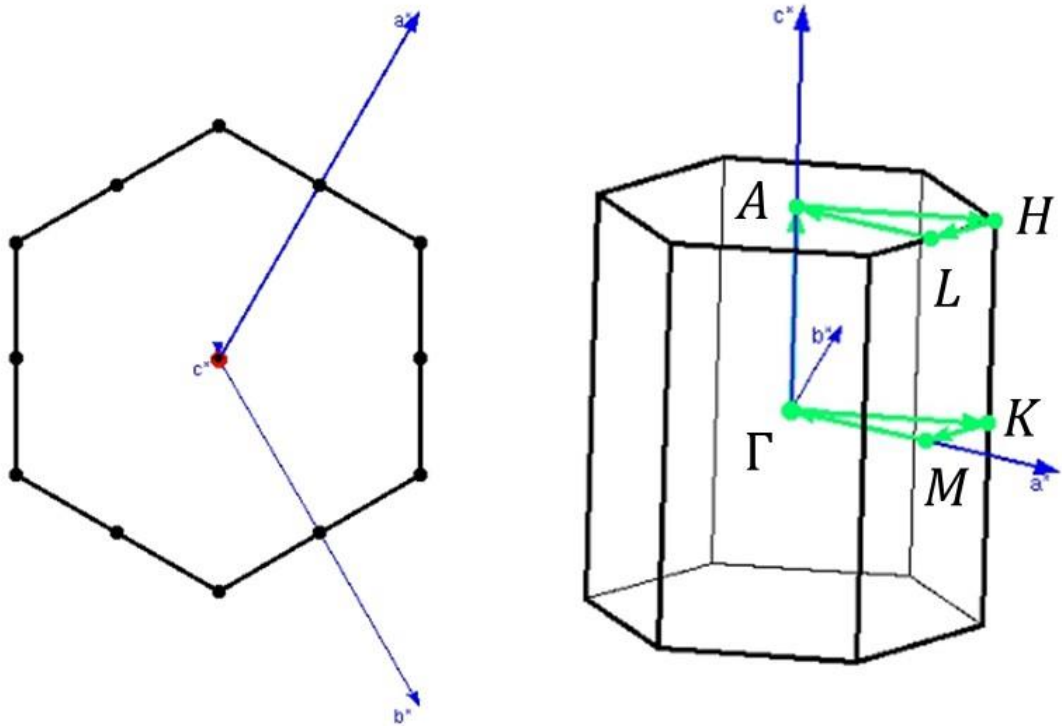

Figure 2 The First Brillouin zone of FeCl$_3$-graphite intercalation compound.

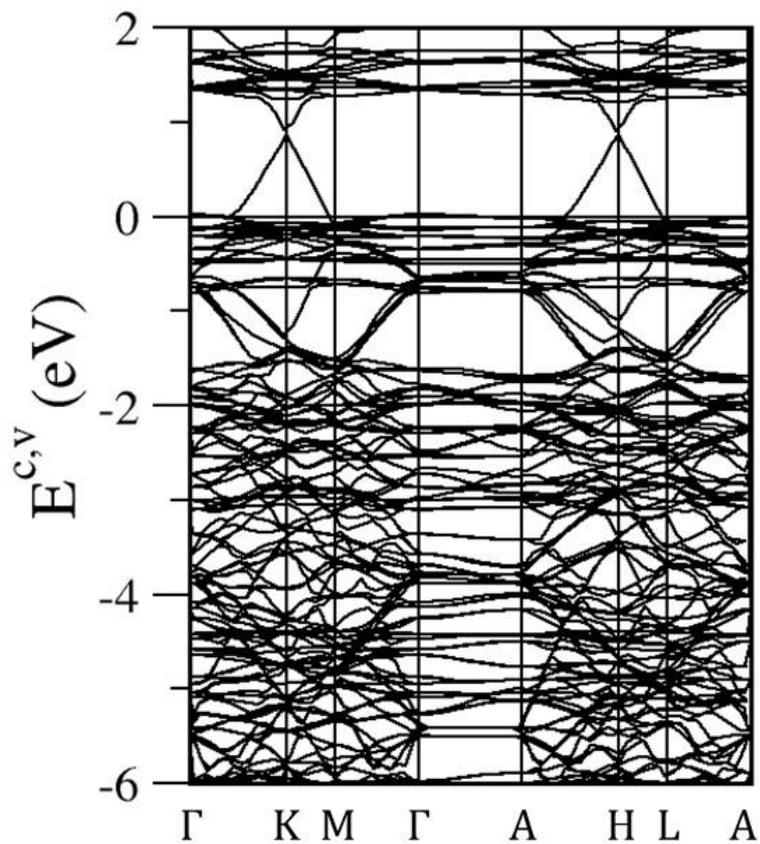

Figure 3 The band structures of FeCl$_3$-grapgite intercalation compound.

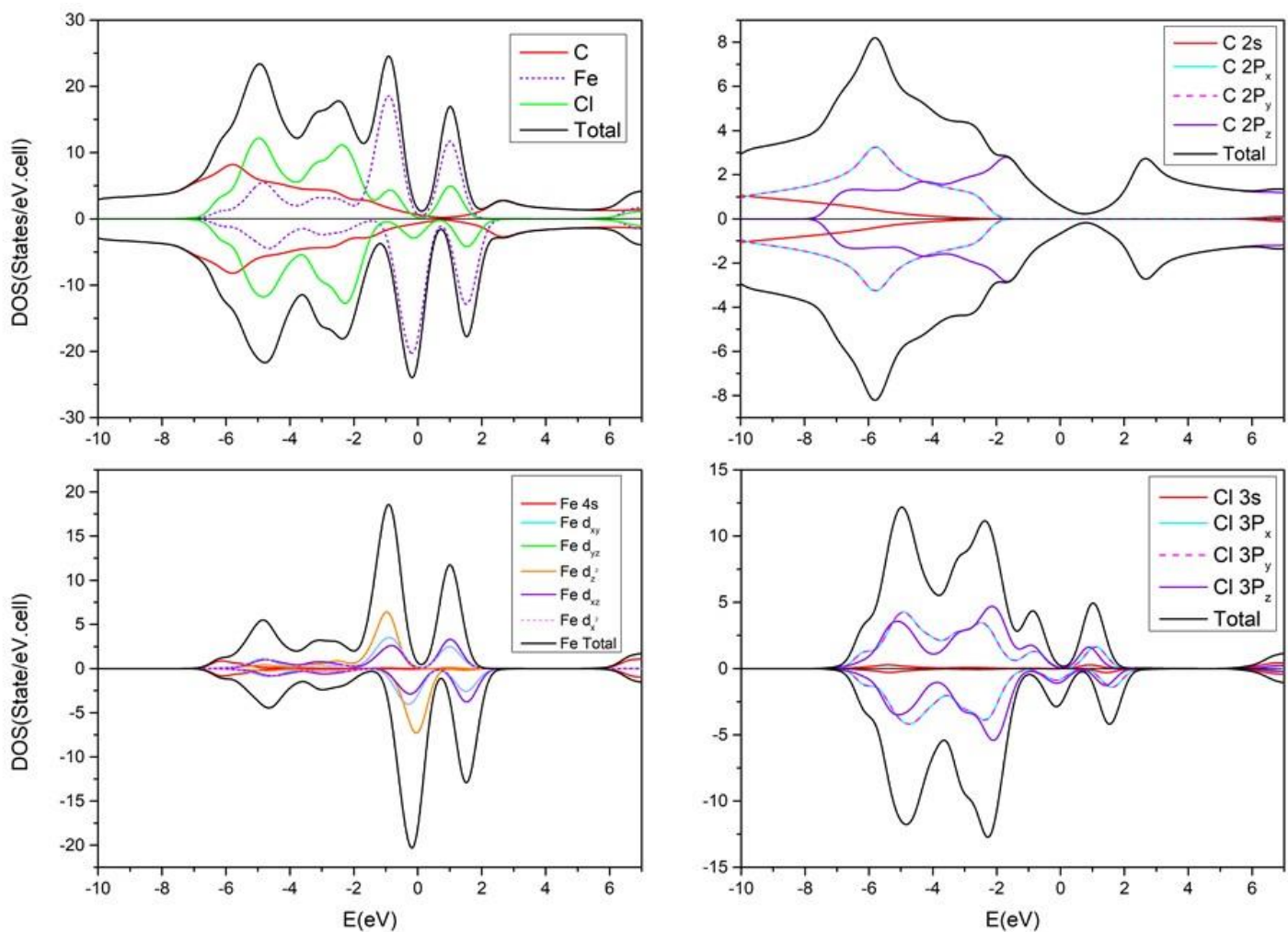

Figure 4 The total density of states and decomposed density of states for C, Fe and Cl, respectively.

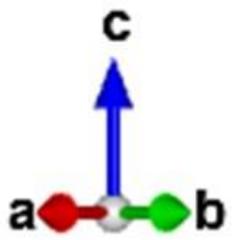
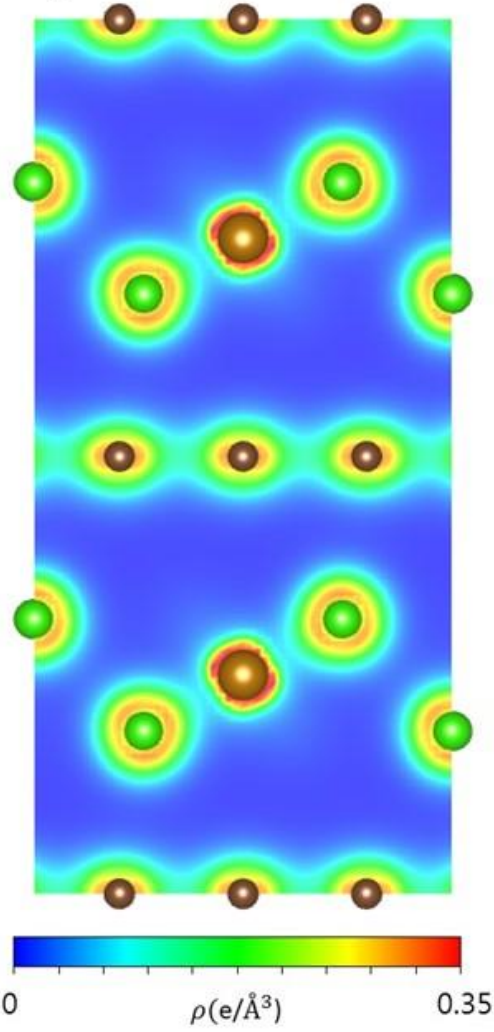
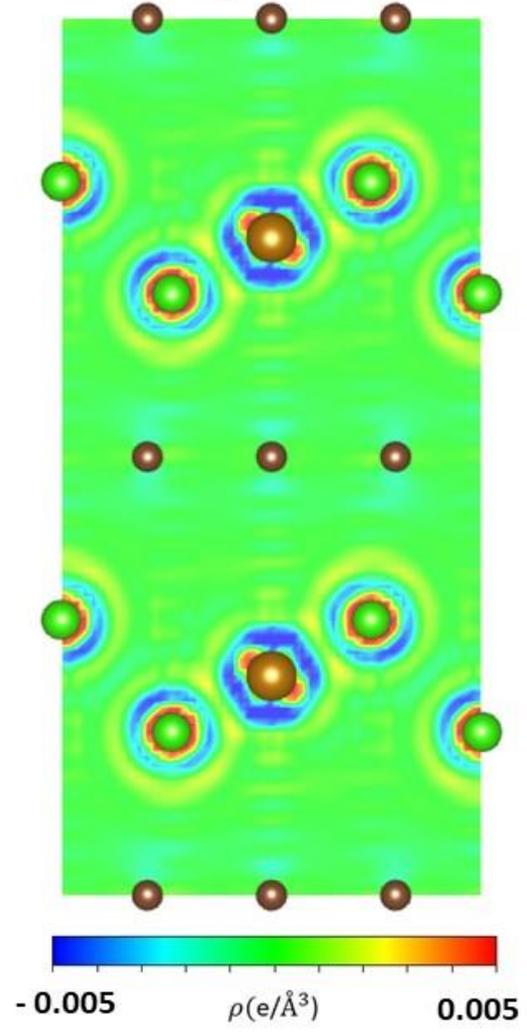

Figure 5 The spatial charge distribution and charge variation.